\documentclass{llncs} 

\usepackage{amsfonts}
\usepackage{amssymb}
\usepackage{amsmath}

\usepackage[figure,table]{totalcount}
\usepackage{graphicx}
\usepackage{url}
\usepackage{color}
\usepackage{parskip}
\usepackage{booktabs}
\usepackage[english]{babel}

\usepackage[ruled]{algorithm}
\usepackage{algorithmic}[1]

\def\R{{\mathbb R}}
\newcommand{\scsc}{\textsf{SC}$^2$ }
\newcommand{\head}[1]{\textnormal{\textbf{#1}}}

\begin{document}

\title{Towards Incremental Cylindrical \\ Algebraic Decomposition in Maple}
\titlerunning{Towards Incremental CAD in Maple} 
\author{Alexander I. Cowen-Rivers\inst{1} and Matthew England\inst{2}}
\authorrunning{A.I. Cowen-Rivers and M. England} 

\institute{
University College London, London, UK \\
\email{a.cowen-rivers.17@ucl.ac.uk}
\and
Coventry University, Coventry, UK \\
\email{Matthew.England@coventry.ac.uk}
}

\maketitle 

\begin{abstract}
Cylindrical Algebraic Decomposition (CAD) is an important tool within computational real algebraic geometry, capable of solving many problems for polynomial systems over the reals.  It has long been studied by the Symbolic Computation community and has found recent interest in the Satisfiability Checking community.    

The present report describes a proof of concept implementation of an Incremental CAD algorithm in \textsc{Maple}, where CADs are built and then refined as additional polynomial constraints are added.   
The aim is to make CAD suitable for use as a theory solver for SMT tools who search for solutions by continually reformulating logical formula and querying whether a logical solution is admissible.

We describe experiments for the proof of concept, which clearly display the computational advantages compared to iterated re-computation.   
In addition, the project implemented this work under the recently verified Lazard projection scheme (with corresponding Lazard valuation).  
\end{abstract}

\newcounter{Count}

\section{Introduction}

We aim to adapt Cylindrical Algebraic Decomposition (CAD) for use with SMT-solvers \cite{BHvMW09}, as part of the \scsc Project which seeks to build collaborations between researchers in Symbolic Computation and Satisfiability Solving \cite{AAB+16a}.  
We report on an implementation of incremental CAD in \textsc{Maple} which can build a CAD and then refine it by incrementally adding polynomials.  The implementation is restricted to Open CAD (full dimensional cells) and the addition of constraints (an SMT solver would also want the ability to remove them).  While a proof of concept implementation,  experiments show clear savings on offer.  

Another minor contribution of the present work is an implementation of the Lazard projection operator (and corresponding valuation for lifting). The operator was proposed in 1994 \cite{lazard_1994}, but shortly after a flaw was found in its proof of correctness (see \cite{MH16} for details).  However, recent work \cite{MPP17} has given an alternative proof (which necessitates some changes to the lifting stage).  It is now the smallest known complete CAD projection operator.

\subsection{Terminology}

We work over $n$-dimension real space $\mathbb{R}^n$ in which there is a variable ordering.

\textbf{Definition \stepcounter{Count}\arabic{Count}} A \textit{decomposition} of the space \(\mathcal{X} \subset \mathbb{R}^n\) is a finite collection of disjoint regions, called \emph{cells}, whose union is \(\mathcal{X}\).

\textbf{Definition \stepcounter{Count}\arabic{Count}} A set is \textit{semi-algebraic} if it can be constructed by finitely many applications of $union$, $intersection$ and $complementation$ operations on sets of the form \(\{ x \in \mathbb{R}^{n}\ |\ \mathbf{f}(x) \geq 0 \}\) where \(\mathbf{f} \in  \mathbb{R}[x_{1},\cdots,x_{n}]\).

\textbf{Definition \stepcounter{Count}\arabic{Count}} A decomposition \(\mathcal{D}\) is \textit{algebraic} if each of its components \( x \in \mathcal{D}\) is a semi-algebraic set.

\textbf{Definition \stepcounter{Count}\arabic{Count}} A finite partition of \(\mathcal{D}\) of \(\mathbb{R}^n\) is called a \textit{cylindrical decomposition} of \(\mathbb{R}^n\) if the projections of any two cells onto any lower dimensional coordinate space with respect to the variable ordering are either equal or disjoint. 

Thus a \textit{cylindrical algebraic decomposition} (CAD) satisfies Definitions $1-4$.

\textbf{Definition \stepcounter{Count}\arabic{Count}} A CAD is \textit{sign-invariant} with respect to a set of input polynomials if each polynomial has a constant sign (positive, negative or zero) on each cell.

CADs may be produced with other invariant properties (see for example \cite{EBD15}, \cite{BDEMW16}) but we assume sign-invariance in the present work.
Each CAD cell is equipped with: a \textit{cell index} which is a list of integers that defines the position of a cell in the decomposition; and a \textit{sample point} of the cell.  The cells we produce also come with a \textit{cell description}: a cylindrical formula, that is, a description of the cell as a sequence of conditions on ordered variables of the form $\ell(x_1, \dots, x_{k})<x_{k+1}<u(x_1, \dots, x_{k})$, where $\ell$ and $u$ may be $\pm\infty$.   

CADs are traditionally produced through a two-stage process:  first projection identifies polynomials of importance for the invariance property and then lifting incrementally builds CADs of $\mathbb{R}^k$ for $k=1, \dots, n$ according to these polynomials.  Decompositions are performed by working at a sample point of a cell, reducing multivariate polynomials to univariate and then decomposing according to the output of real root isolation.  For a fuller introduction see the lecture notes \cite{jirstrand_1995}.

\subsection{Example}

\begin{figure}[H]
  \centering
  \begin{minipage}[b]{0.48\textwidth}
      \includegraphics[width=.95\linewidth]{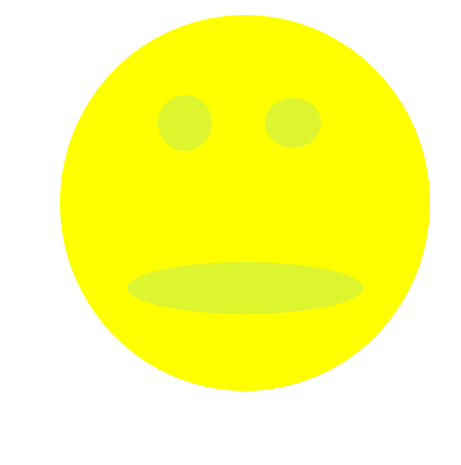}
    \caption{Gingerbread face formed by 4 bivariate polynomials}
    \label{fig:ging}
  \end{minipage}
  \hfill
  \begin{minipage}[b]{0.48\textwidth}
      \includegraphics[width=.95\linewidth]{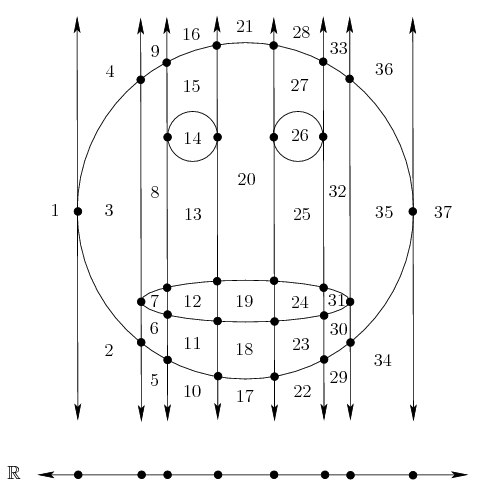}
    \caption{A CAD of Fig \ref{fig:ging}.  Numbers correspond to open (full dimensional) cells.}
        \label{fig:ging2}
  \end{minipage}
\end{figure}

We give a visual example\footnote{Inspired by \url{http://planning.cs.uiuc.edu/node296.html}}.  The gingerbread face in Figure \ref{fig:ging} is formed by four closed curves, each of which defined by a bi-variate polynomial equation.  A corresponding sign-invariant CAD of $\mathbb{R}^2$ is visualised in Figure \ref{fig:ging2}.
We label the $37$ open cells (those of two dimensions).  There are a further $28$ partially open (1-dimensional line segments) and $28$ closed cells (isolated points) giving $93$ CAD cells in total.
Of course, in many industrial and SMT applications the polynomials will not form such aesthetically pleasing geometric shapes.

\subsection{Report plan}

We aim to work with CADs that change incrementally by constraint.  I.e. given a CAD of $\mathbb{R}^n$ sign-invariant for polynomials $\{f_1, \dots, f_m\}$ we aim to adapt this to one sign-invariant for $\{f_1, \dots, f_m, f_{m+1}\}$ or $\{f_1, \dots, f_{m-1}\}$.  The present report deals only with the first problem.  Such incrementality is needed to use CAD in SMT, and could also benefit CAD directly as a way of reducing the search space.
We proceed by considering the changes required in Projection (Section 2) and Lifting (Section 3) alongside the issues in using the Lazard projection operator.  We finish with a summary and plans for future work in Section 4.
A larger report with additional details we do not have room for here is available on \textsc{arXiv} \cite{SummerReport}. 

\section{Projection}

\subsection{Lazard projection}

The present project built upon code from the \textsc{ProjectionCAD} package \cite{EWBD14} for \textsc{Maple}.  This implemented the McCallum family of projection operators \cite{McCallum1998}, \cite{BDEMW16} and our first step was to adapt this for the Lazard projection scheme.
All projection operators take a set of polynomials and produce another set in one less variable.  The Lazard operator is essentially a subset of the McCallum operator.  Both take discriminants and cross-resultants of the input polynomials.  The Lazard operator then takes in addition leading and trailing coefficients while the McCallum operator takes all coefficients.  Thus adaptations to the projection algorithms were fairly minimal here (see \cite{SummerReport} for algorithms).  The main computational differences occur in the lifting stage as discussed later.

\newpage

\subsection{Worked example}

We describe a worked example to illustrate what projection does and to use later to illustrate incremental projection.  We work with a system of two polynomials $\textbf{F}_1$ and seek a sign-invariant CAD:
\begin{equation}
\label{eq:f1}
\textbf{F}_1=\{\underbrace{x_1^2+x_2^2-1}_{f_1},\underbrace{x_1^3-x_2^2}_{f_2}\}.
\end{equation}

Since the problem involves only two variables, we need only a single projection (which we do with respect to $x_2$).  The leading coefficients are constant, but the trailing coefficients clearly identify the points on the $x_1$ line at $0$ and $\pm 1$.  The discriminants do not identify anything more but the resultant
\begin{equation}
\label{eq:resEx}
\texttt{Resultant}(f_1,f_2,x_2) = (x_1^3 + x_1^2 - 1)^2 
\end{equation}
identifies $\pm \alpha_1 = \approx \pm 0.7549$\footnote{We give a decimal approximation but emphasise that CAD would use the full algebraic number representation: that $\alpha_1$ is the sole real root of (\ref{eq:resEx}) in $(0,1)$.}.
Thus the real line is decomposed into 11 cells according to these 5 points.

Figure \ref{fig:we1} plots the graphs of these functions along with the real roots isolated.  We see they mostly correspond to geometrically relevant features ($-\alpha_1$ corresponds to an intersect in $\mathbb{C}^2$).

\subsection{Incremental Lazard projection}

Our second step was to adapt the Lazard Projection algorithms to calculate new projection polynomials incrementally. 

We continue our example to illustrate the incremental working.  Suppose we take $\textbf{F}_1$ from above and also the polynomial forming a line, $f_3=x_2-x_1$ to give
\begin{equation}
\label{eq:4}
\textbf{F}_2=\{\underbrace{x_1^2+x_2^2-1,x_1^3-x_2^2}_{\textbf{F}_1},\underbrace{x_2-x_1}_{f_3}\}.
\end{equation}
We will compute all the projection polynomials discussed above and some additional ones. The discriminant and leading coefficient of $f_3$ are constant, and the trailing coefficient identifies $x_1=0$ which was already present from the trailing coefficient of $f_2$.  Similarly, the resultant of $f_2$ and $f_3$ is $x_1^2(x_1-1)$ identifies two more roots we saw already.  However $\texttt{Resultant}(f_1,f_3,x_2) = 2x_1^2-1$ identifies two new points, $\pm \alpha_2 = \pm 1/\sqrt{2} \approx \pm 0.7071$.

\begin{figure}[H]
  \centering
  \begin{minipage}[b]{0.47\textwidth}
    \includegraphics[width=\textwidth]{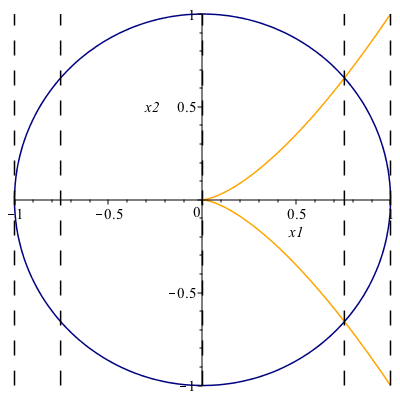}
    \caption{The blue curve is $f_1$ and the orange $f_2$.  Dotted lines show the projection roots $\in \mathcal{R}$.
    \label{fig:we1}}
  \end{minipage}
  \hfill
  \begin{minipage}[b]{0.47\textwidth}
    \includegraphics[width=\textwidth]{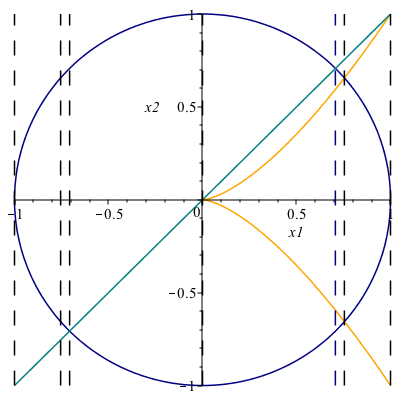}
    \caption{As Fig. 3 but with the additional teal curve $f_2$ and the additional roots identified.
    \label{fig:we2}}
  \end{minipage}
\end{figure}

Figure \ref{fig:we2} shows that the two new roots correspond to the two new intersections of the straight line with the circle.  

We developed Algorithms \ref{alg:5} and \ref{alg:4} to implement such an incremental projection.  The pseudo code is closely linked to the \textsc{Maple} implementation.  The former computes all the projection polynomials via calls to the latter which performs one projection.
The adjustments (highlighted in blue) required to make the original projection code (see \cite{SummerReport}) incremental were: 
\begin{enumerate} 
\item To output from \texttt{ProjectionPolysAdd} the full table of projection polynomials organised by the main variable and re-input this with incremental calls.  
%
\item The process and pass polynomials separately into \texttt{ProjectionAdd}.
%
\item To calculate only the new projection polynomials and store appropriately.
\end{enumerate}

\begin{algorithm}[t]
\caption{ProjectionPolysAdd}
\label{alg:5}
{\fontsize{10}{15}\selectfont
\begin{algorithmic}[1]
\STATE \textbf{Input:} Set of calculated projection polynomials $prev \in \mathbb{R}[x_n,\ldots,x_{1}]$ and set of new polynomials $new\in \mathbb{R}[x_{n},\ldots,x_1]$ 
\textcolor{blue}{(Adjustment 1)}
\STATE \textbf{Output:} All projection polynomials $\in \mathbb{R}[x_{n}, \dots, x_{1}] $
\STATE \textbf{Procedure} \texttt{ProjectionPolysAdd}
(Compute all projection polynomials)
\STATE $dim \leftarrow$ Number of variables $+1$
\STATE $pset[0] = table()$ 
\STATE $pset[0] \leftarrow$ Primitive set from \textcolor{blue}{$new$}, wrt variable $x_n$ \textcolor{blue}{(Adjustment 2)}
\STATE $pset[0] \leftarrow$ Square free basis set from $pset[0]$, wrt variable $x_n$ 
\STATE $pset[0] \leftarrow$ Set of factors from $pset[0]$, wrt variable $x_n$ 
\STATE $cont \leftarrow$ Set of contents of \textcolor{blue}{$new$}, wrt $x_n$ \textcolor{blue}{(Adjustment 2)}
\FOR{\texttt{i from 1 to dim-1}}
\STATE $out \leftarrow$ \texttt{ProjectionAdd}$(\textcolor{blue}{prev[i-1]},pset[i-1])$ \textcolor{blue}{(Adjustment 2)}
\STATE $pset[i] \leftarrow (out \ \cup \ cont)$
\STATE $cont \leftarrow$ Content set from $pset[i]$, wrt variable x$_{n-i}$ 
\STATE $pset[i] \leftarrow$ Prime set from $pset[i]$, wrt variable $x_{n-i}$
\STATE $pset[i] \leftarrow$ Square free basis from set $pset[i]$, wrt variable $x_{n-i}$ 
\STATE $pset[i] \leftarrow$ Set factors from $pset[i]$, wrt variable $x_{n-i}$ 
\ENDFOR

\STATE $pset[dim-1]    \leftarrow$ Remove constant multiples from $pset[[dim-1]$
\STATE $ret \leftarrow pset[[dim-1]$ 

\STATE \textbf{return} \textcolor{blue}{{$pset$} (Adjustment 1)}
\end{algorithmic}
}
\end{algorithm}

\begin{algorithm}[p]
\caption{ProjectionAdd}
\label{alg:4}
{\fontsize{10}{15}\selectfont
\begin{algorithmic}[1]
\STATE \textbf{Input:} Sets of polynomials $new=\{f_1,\dots,f_m\} \in \mathbb{R}[x_n,\ldots,x_1]$, \textcolor{blue}{$old \in \mathbb{R}[x_{n-1},\ldots,x_{1}]$}, and variable ordering 
\textcolor{blue}{(Adjustment 2)}
\STATE \textbf{Output:} Set of polynomials $Pset=\{p_1,\ldots,p_q\} \in \mathbb{R}[x_{n-1},\ldots,x_{1}] $
\STATE \textbf{Procedure} \texttt{ProjectionAdd}
\STATE $Polys \leftarrow$ Primitive set from \textcolor{blue}{$new$}, wrt variable $x_n$ \textcolor{blue}{(Adjustment 2)}
\STATE $Cont \leftarrow$ Content set from \textcolor{blue}{$new$}, wrt variable $x_n$ \textcolor{blue}{(Adjustment 2)}
\STATE $Polys \leftarrow$ Square free basis set from $Polys$, wrt variable $x_n$ 
\STATE $Pset1 = table():$
\FOR{\texttt{i from 1 to number of elements of $Polys$}}
        \STATE $Pol\leftarrow Polys[i] $
        \STATE $clist \leftarrow $ Lazard coefficient set from $Pol$, wrt to $x_n$
        \STATE $temp \leftarrow$ Discriminant set from $Pol$, wrt $x_n$ 
        \STATE $temp \leftarrow$ Remove constant multiples from $temp$ 
        \STATE $Pset1[i] \leftarrow$ union $temp$ \& $clist$ 
\ENDFOR
\STATE $Pset2 = table():$
\FOR{\texttt{i from 1 to number of $Polys$}}
\FOR{\texttt{j from i+1 to number of $Polys$}}
    \STATE     $Pset2[i,j] \leftarrow$ Resultant of $Polys[i]$ and $Polys[j]$, wrt to variable $var$
    \STATE     $Pset2[i,j] \leftarrow$ Remove constant multiples from $Pset2[i,j]$
\ENDFOR
\ENDFOR \textcolor{blue}{
\STATE $oldset \leftarrow old$ 
\STATE $Pset3 = table():$
\FOR{\texttt{i from 1 to number of $Polys$}}
\FOR{\texttt{j from 1 to size$(oldset)$}}
    \STATE     $Pset3[i,j] \leftarrow $ Resultant of $Polys[i]$ and $oldset[j]$, wrt to variable $var$
    \STATE     $Pset3[i,j] \leftarrow $ Remove constant multiples from $Pset3[i,j]$
\ENDFOR
\ENDFOR (Adjustment 3)}
\STATE $Pset \leftarrow$ union $(cont,Pset1,Pset2,\textcolor{blue}{Pset3})$  
\STATE $Pset \leftarrow$ Remove constant multiples from $Pset$ 
\STATE \textbf{return} $Pset$
\end{algorithmic}
}
\end{algorithm}

\subsection{Incremental projection experimental results}

There were seven elements of the projection set to calculate in the original projection system \texttt{ProjL($F_1$)} and after the addition of the extra polynomial \texttt{ProjL($F_2)$} had 12.  By avoiding full recomputation, we had to compute only the extra five. 
At the cell level the savings are more significant: on the real line there were 5 original roots (11 cells), and two new ones (making 15 cells).

We performed experiments to see how such savings transferred into computation time.  We created examples using the random polynomial function \texttt{randpoly} in \textsc{Maple}. 
Testing was conducted through an external bash script creating new \textsc{Maple} instances to avoid any result caching.  The testing code is  available online\footnote{\url{https://github.com/acr42/InCAD.git}}.  

\subsubsection{Bivariate polynomials} 

We created 60 pairs of bivariate polynomials, considered first finding the projection of one and then incrementing the projection by including the other.  On average it was 16\% faster to increment compared to computing the projection for both polynomials together.  However, there was a large variance: the cases which were faster were on average 55\% faster, while a small number of cases were slower, by as much as $87\%$.

\begin{center}
\begin{tabular}{@{}*4l@{}}
  \toprule[1.5pt]
  \multicolumn{1}{c}{\head{Projection}} &
    \multicolumn{2}{c}{\head{Results}} \\ 
  \head{} & \head{Classical} & \head{Incremental} & \head{}\\
  \cmidrule(l){1-1}\cmidrule(r){2-4}
  \verb|Variance| & \rmfamily 0.0004660s & \rmfamily  0.0006425s & \rmfamily  \textcolor{red}{\textbf{27.46\%}} Larger\\
  \verb|Mean| & \rmfamily 0.03743s & \rmfamily  0.0315s & \rmfamily  \textcolor{green}{\textbf{15.85\%}} Faster \\
  \verb|Lower Quartile| & \rmfamily 0.024s & \rmfamily  0.008s & \rmfamily  \textcolor{green}{\textbf{66.66\%}} Faster \\
  \verb|Median| & \rmfamily 0.0285s & \rmfamily  0.015s & \rmfamily  \textcolor{green}{\textbf{47.39\%}} Faster\\
  \verb|Upper Quartile| & \rmfamily 0.03675s & \rmfamily  0.05525s & \rmfamily  \textcolor{red}{\textbf{50.34\%}} Slower \\
  \bottomrule[1.5pt]
\end{tabular}
\end{center}

\subsubsection{Trivariate polynomials}

We next created 80 further examples with pairs of trivariate polynomials, in this case restricting to 4 terms per polynomial.  Here there were greater savings, on average 29\% faster to increment than to compute altogether.  There was also a smaller variance in the timings of the example set, although it was still the case that a few examples were slower to increment than recompute.

\begin{center}
\begin{tabular}{@{}*4l@{}}
  \toprule[1.5pt]
  \multicolumn{1}{c}{\head{Projection}} &
    \multicolumn{2}{c}{\head{Results}} \\ 
  \head{} & \head{Classical} & \head{Incremental} & \head{}\\
  \cmidrule(l){1-1}\cmidrule(r){2-4}
  \verb|Variance| & \rmfamily 0.002743s & \rmfamily  0.002205s & \rmfamily  \textcolor{green}{\textbf{24.39\%}} Smaller\\
  \verb|Mean| & \rmfamily 0.06739s & \rmfamily  0.04809s & \rmfamily  \textcolor{green}{\textbf{28.64\%}} Faster \\
  \verb|Lower Quartile| & \rmfamily 0.02475s & \rmfamily  0.013s & \rmfamily  \textcolor{green}{\textbf{47.47\%}} Faster \\
  \verb|Median| & \rmfamily 0.0625s & \rmfamily  0.035s & \rmfamily  \textcolor{green}{\textbf{44.00\%}} Faster\\
  \verb|Upper Quartile| & \rmfamily 0.09425s & \rmfamily  0.07525s & \rmfamily  \textcolor{green}{\textbf{20.16\%}} Faster \\
  \bottomrule[1.5pt]
\end{tabular}
\end{center}

We suggest that the extra overheads of the incremental approach will become less important in comparison to the savings as the number of variables increase: indeed, this follows from the well-known complexity results on CAD. 

\section{Lifting}

\subsection{Lifting after Lazard projection}

We had to make changes to the lifting code in \textsc{ProjectionCAD}, not just to allow for incrementality but also to validate the use of the Lazard projection operator \cite{MPP17}.  The McCallum projection operator \cite{McCallum1998} is known to be incomplete if it occurs that a projection polynomial is nullified over the sample point of a cell.  For example, the polynomial $(y^2-2)w+z(y-x^2+x+2)$ is nullified over a cell in $(x,y,z)$-space with sample point $(\sqrt{2},-\sqrt{2},1)$.  When lifting after McCallum projection, one must check for this situation and warn the user that above the cell in question we are not guaranteed sign-invariance.

With the Lazard operator (as proved valid in \cite{MPP17}) we can avoid such checks and warnings but we must do some additional work during lifting to recover information lost by nullification, as outlined in Algorithm \ref{alg:7}.  With the previous example we would first substitute for $x = \sqrt{2}$ to get 
$(y^2-2)w+z(y+\sqrt{2})$; but then before substituting for $y$ we must divide by $y + \sqrt{2}$ to give $(y-\sqrt{2})w+z$.  We can only then substitute for $y$ to give $-2\sqrt{2}w + z$ and finally $z$ to give $-2\sqrt{2}w + 1$.  We thus must lift with respect to this univariate polynomial in $w$, creating necessary cell divisions that would have been lost by nullification.

This process is difficult when involving irrational sample points.  However, since our prototype implementation lifts only over open cells, we have avoided such difficulties for now. Our implementation produces Open CADs, avoiding costly algebraic number calculations, but still getting a good understanding of the solution set.

\newpage

\textbf{Definition \stepcounter{Count}\arabic{Count}} An \textbf{Open-CAD} is produced by lifting over open intervals only\footnote{Not actually a decomposition of $\R^n$ as missing boundaries of the $n$-dimensional cells.}. 

\begin{algorithm}[ht]
\caption{Lazard valuation}
\label{alg:7}
{\fontsize{10}{15}\selectfont
\begin{algorithmic}[1]
\STATE \textbf{Input:} A polynomial $f \in \mathbb{R}[x_1,\ldots,x_d]$, and $\texttt{sp}=[r_1,\ldots,r_{d-1}] \in \mathbb{R}^{d-1}$.
\STATE \textbf{Output:} List of roots. 
\STATE \textbf{Procedure} \texttt{LazardValuation}:
\STATE Set $\texttt{roots}$ to be an empty list 
\FOR {\texttt{$j$ from $1$ to $d-1$}}
\WHILE {\texttt{$f(r_1, \dots, r_j)=0$}}
\STATE $f \leftarrow f/(x_j-r_j)$ 
\STATE substitute $x_j = r_j$ into $f$.
\ENDWHILE
\ENDFOR 
\STATE \textbf{return} f:
\end{algorithmic}
}
\end{algorithm}

We will now discuss our approach for incremental lifting, which can be thought of graphically as a form of acyclic tree merge, as shown later.  

\subsection{Worked example}

We will describe lifting the projection polynomial system defined previously \texttt{ProjL}($\textbf{F}_1$). 
Recall that we identified four points on the real line: 
$\{-1,0,\alpha_1,1\}$.
Thus, we need to choose a sample value from the 9 cells in the decomposition:
\begin{equation}
\label{ineq:1}
  \begin{array}{lll}
     a_1=\{ x_1<-1\}, &a_2=\{x_1=-1\}, &a_3=\{-1<x_1<0\},\\
     a_4=\{x_1=0\},&a_5=\{0<x_1<\alpha_1\}, &a_6=\{x_1=\alpha_1\},\\
     a_7=\{\alpha_1<x_1<1\}, &a_8=\{x_1=1\},&a_9= \{1<x_1\} 
    \end{array}
\end{equation}
We are forced to pick non-rational sample points for cell $a_6$ but the others can be rational.  We choose sample points: 
$
(-2, -1, -\tfrac{1}{2},0,\tfrac{1}{2},\alpha_1,\tfrac{9}{10}, 1, 2).
$

\begin{figure}[t]
  \centering
  \begin{minipage}[b]{\textwidth}
    \includegraphics[width=\textwidth]{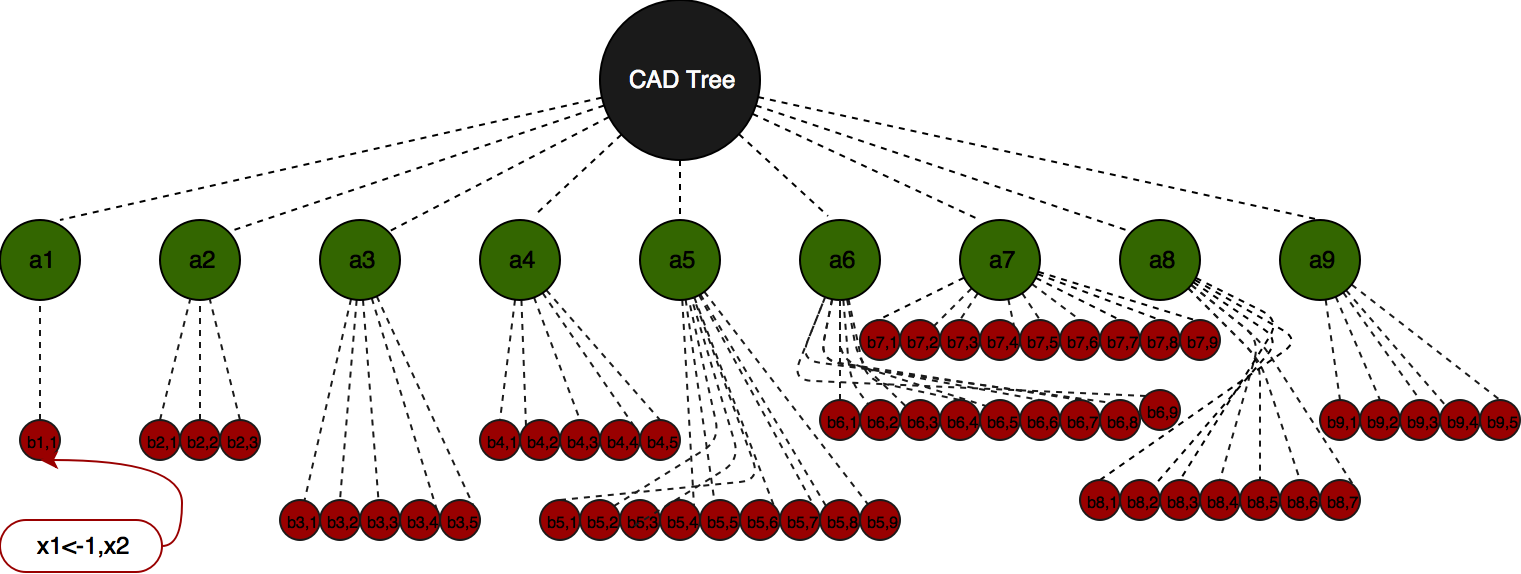}
    \caption{CAD tree of $F_1$. Green nodes are in the first dimension and red the second.}
    \label{tree:1}
  \end{minipage} 
\end{figure}

We lift over each cell at the designated sample point by isolating real roots of the univariate polynomials in $x_2$ we get by from the Lazard valuation at the sample point in $x_1$.  
Below, $p_{i,j}$ denotes the polynomial acquired after applying the Lazard valuation method to the $i$'th sample point, on the $j$'th polynomial from $F_1$.  For example, if we use sample point $-\tfrac{1}{2}$ for cell $a_3$ then polynomial $f_1$ becomes $\tfrac{1}{4} + x_2^2 - 1$ which has two real roots at $\pm \beta_0 = \pm \sqrt{3}/2 \approx \pm 0.8660$.
We proceed this way to generate our CAD cells in $\R^2$.  The structure is as shown in Figure \ref{tree:1}.  For a full list of the new cell descriptions see \cite{SummerReport}.

\subsection{Incremental Lazard lifting}

The general concept of how we solved this stage of the problem was to think of it as solving a graph (tree) attachment/detachment problem. One should think of the old CAD as having a tree structure which we save: where nodes are cells; and branches link a cell to its parent (cell it projects onto), or child (decomposition in a cylinder above) cells. At each depth of the CAD/tree, say depth $p$, are all the cells within $\mathbb{R}^{p}$ before we lifted to $\mathbb{R}^{p+1}$. We  go through a worked example.

We perform an incremental lift on the polynomial system $\textbf{F}_1$, incremented by a new polynomial $f_4=x_1^3+x_2^2$, forming the new system (\ref{eq:f3}). 
\begin{equation}
    \textbf{F}_3=\{\underbrace{x_1^2+x_2^2-1,x_1^3-x_2^2}_{\textbf{F}_1},\underbrace{x_1^3+x_2^2}_{f_4}\} 
    \label{eq:f3}
\end{equation}
The new system is symmetrical about the $y$ axis as you can see in Figure 8. 

\begin{figure}[H]
  \centering
  \begin{minipage}[b]{0.41\textwidth}
    \includegraphics[width=\textwidth]{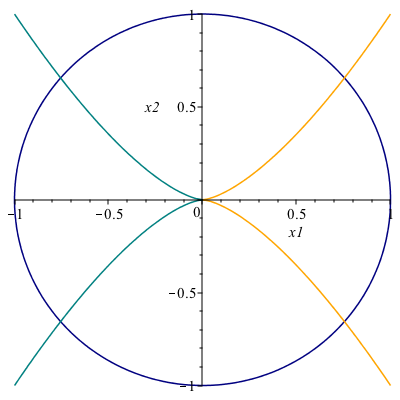}
    \caption{
    The blue curve is $f_1$, the orange $f_2$ and the teal $f_4$.}
  \end{minipage}
  \hfill
  \begin{minipage}[b]{0.41\textwidth}
    \includegraphics[width=\textwidth]{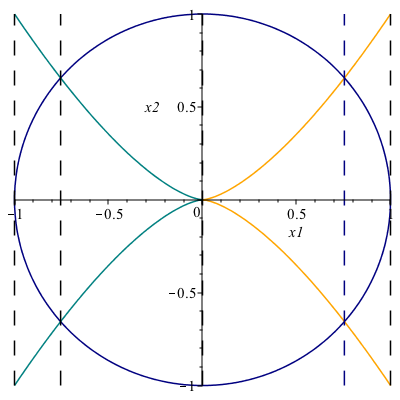}
    \caption{Dotted lines show the projection roots.}
  \end{minipage}
\end{figure}

We skip the projection steps\footnote{See "Worked Examples" worksheet in:  \url{https://github.com/acr42/InCAD.git}}.  
The addition polynomials identify one further point on the real line: $-\alpha_1$.  
The decomposition of the real line is now:
\begin{equation*}
  \begin{array}{l}
  \label{ineq:inlift}
     a_1=\{ x_1<-1\}, a_2=\{x_1=-1\},\textcolor{blue}{ a_3=\{-1<x_1<-\alpha_1\}},\\
    \textcolor{blue}{a_4=\{x_1=-\alpha_1\}},\textcolor{blue}{a_5=\{-\alpha_1<x_1<0\}}, a_6=\{x_1=0\},
     a_7=\{0<x_1<\alpha_1\}, \\ a_8=\{x_1=\alpha_1\},a_9= \{\alpha_1<x_1<1\},a_{10}= \{x_1=1\},a_{11}= \{1<x_1\} \\
     \end{array}
\end{equation*}
and our sample points are:
$-2,-1,\textcolor{blue}{-\tfrac{9}{10}},\textcolor{blue}{-\alpha_1},-\tfrac{1}{2},0,\tfrac{1}{2},\alpha_1,\tfrac{9}{10},1,2$.

We first test whether each sample point leads to new roots when lifting with $f_4$. If so we must refine the decomposition in the cylinder above.
%
For example, on $a_2$ we have sample point $x_1=-1$, and the Lazard valuation of $f_4$ is $x_2^2-1$ with real roots at $\pm 1$.  However, we had already identified these from other polynomials, so no change is required.  However, on $a_5$ with sample point $-\tfrac{1}{2}$, $f_4$ evaluates to $x_2^2 - \tfrac{1}{8}$ and we find two new real roots.  We thus refine the decomposition above.

We then lift over the new sample points with respect to all polynomials creating new decompositions. For example, at $x_1=\tfrac{9}{10}$ we have two roots from the valuation of $f_1$ and another two from $f_4$ (so decomposition above into 9 cells).
Figures 10-12 show the new CAD tree structure and its split into new and unchanged cells, illustrating potential savings.  Full cell descriptions are in \cite{SummerReport}.  

\begin{figure}[t]
  \centering
  \begin{minipage}[b]{0.9\textwidth}
    \includegraphics[width=\textwidth]{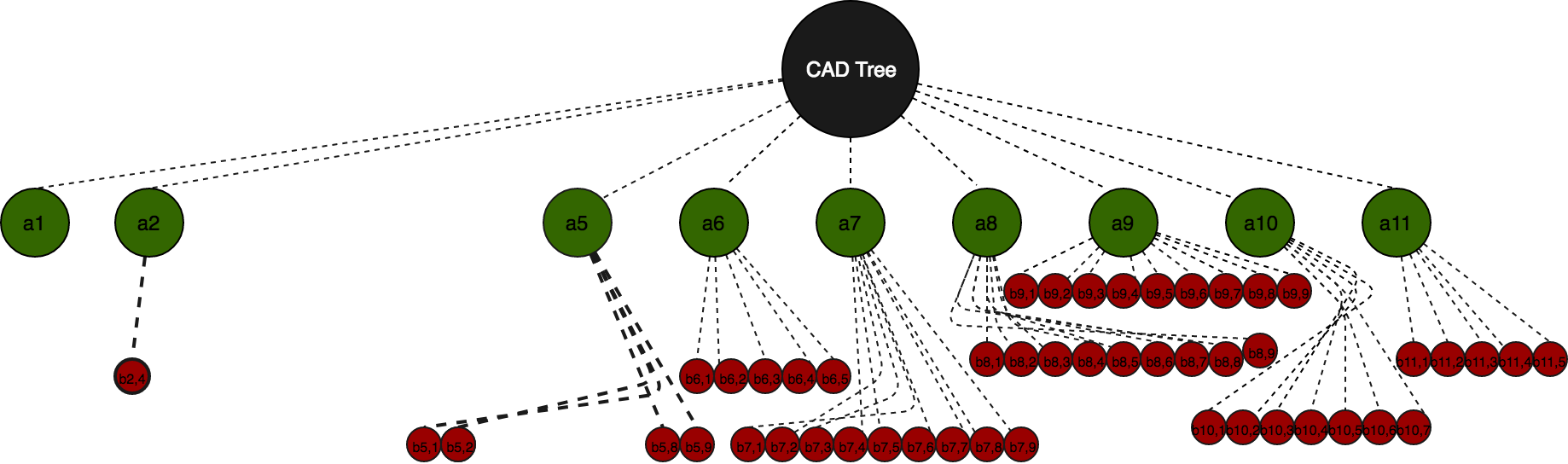}
    \caption{CAD tree of \textbf{unchanged cells} from $F_1$ incremented by $f_4$.
    }
  \end{minipage} 
\end{figure}

\begin{figure}[t]
  \centering
  \begin{minipage}[b]{0.9\textwidth}
    \includegraphics[width=0.7\textwidth]{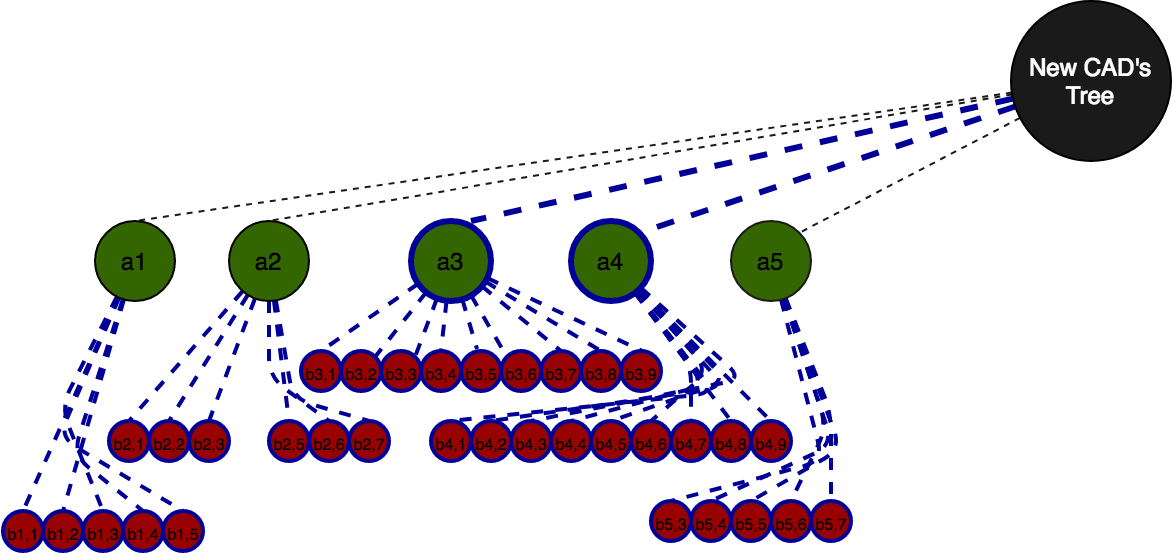}
    \caption{CAD tree of \textbf{new cells} from $F_1$ incremented by $f_4$. Blue outlines around lines/ nodes represent new connections / cells.}
  \end{minipage} 
\end{figure}

\newpage

\begin{figure}[t]
  \centering
  \begin{minipage}[b]{0.9\textwidth}
    \includegraphics[width=\textwidth]{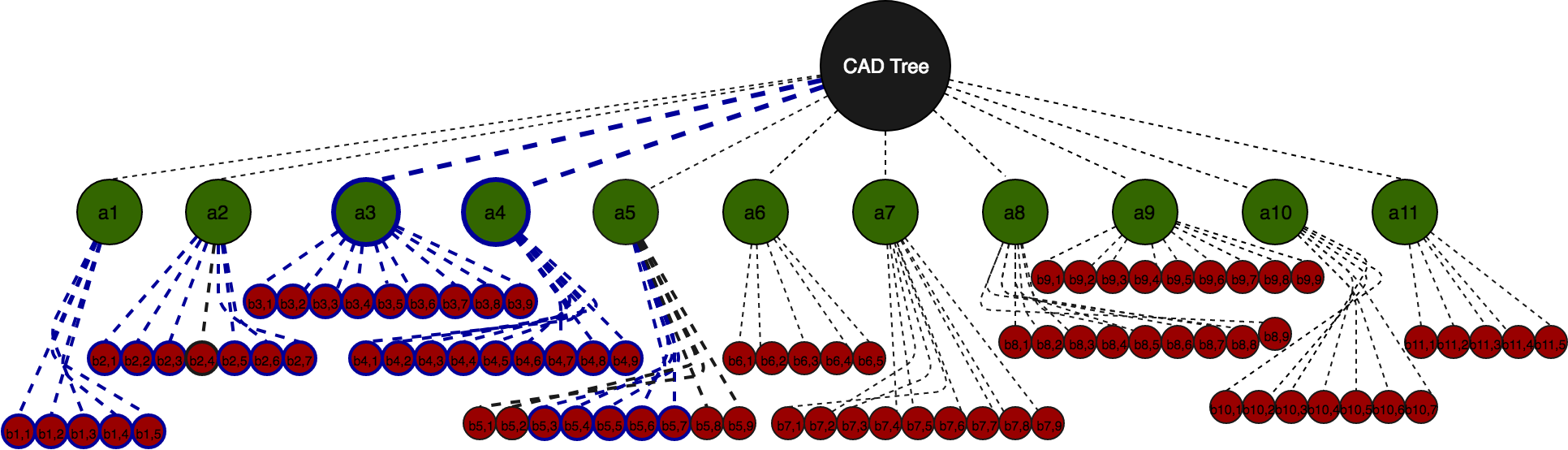}
    \caption{CAD tree of $F_1$ incremented by $f_4$: a merger of Fig. 10 and 11.
    }
  \end{minipage} 
\end{figure}

\subsection{Algorithms}

The two new algorithms created for incremental lifting were \texttt{LiftSetupAdd} (Algorithm \ref{alg:10}) concentrating on the base phase (CAD of the real line) and \texttt{LiftAdd} (Algorithm \ref{alg:liftadd}) which deals with the rest of the CAD. 
The non-incremental versions can be found in the report  \cite{SummerReport}. 

When incrementing the lift stage, we can think of it as starting at the root node of the old CAD tree and working our way down it, one depth level at a time, until we reach the leaves. In the process of working down the old tree, we will be creating subtrees, which will later be reconnected to the unchanged tree, to form the new incremented CAD tree. 
\par 

One tree (\texttt{UnchangedCells}) is a strict subset of the old trees nodes and edges, discovered through analysing the old structure down and Lazard valuating on each cell not marked as \emph{new}, at each depth $p$ with new projection polynomials in $R^{p+1}$. Then, if there are new real roots discovered, we prune inspected cells children, and the cell is sent to the \texttt{NewCells} set for full re-computation. In the \texttt{NewCells} set, we will then use this cell to form a subtree, to later be reconnected with the \texttt{UnchangedCells} tree. Each cell in the \texttt{NewCells} list is a subtree, to later be reconnected via source indices. 

When going through the old CAD tree structure, we have only two cases: 

$\textbf{CASE 1:}$ When a node has new children: \\
New real roots have been acquired from one of the new projection polynomials. We start by pruning all of the child branches in the old tree structure, by labelling them as \emph{new}, then performing a full lift onto the set of all projection polynomials from $\mathbb{R}^{k}, \ldots, \mathbb{R}^{n}$, where $k$ is the depth the new root was discovered. 
When we label a cell as \emph{new}, effectively that halts tree growth in the \texttt{UnchangedCells} structure, so that later on we can attach the branch extensions, gained from the incremental lift.
Such cells 
have an updated source index, as its source cell would now be saved in a new tree structure with a new index. 

$\textbf{CASE 2:}$ When a node has no new children. \\
The \emph{new} flag is passed down to children from parents.  Cells which are not \emph{new} are stored in \texttt{UnchangedCells}. 
Cells here only have Lazard valuation at the new projection polynomial (rather than all projection polynomials). If this leads to a new root, we move it over to the \texttt{NewCells} structure. Otherwise, we continue with its child cells.

When moving cells into \texttt{UnchangedCells}, we make sure that the indexing of each cell does not clash with that of the indexing in \texttt{NewCells} at each depth level in the tree. We then merge the \texttt{NewCells} and \texttt{UnchangedCells} trees, forming the full incremented CAD tree. At each stage of the lift, we merge-sort the list.

\subsection{Incremental lifting experimental results}

We conducted testing for the incremental lift method on the same examples used to test the incremental projection earlier.

\textbf{Bivariate polynomials}
On average 30\% faster than recomputing.

\begin{center}
\begin{tabular}{@{}*4l@{}}
  \toprule[1.5pt]
  \multicolumn{1}{c}{\head{Lift}} &
    \multicolumn{2}{c}{\head{Results}} \\ 
  \head{} & \head{Classical} & \head{Incremental} & \head{}\\
  \cmidrule(l){1-1}\cmidrule(r){2-4}
  \verb|Variance| & \rmfamily 0.003734s & \rmfamily  0.002903s & \rmfamily  \textcolor{green}{\textbf{28.63\%}} Smaller\\
  \verb|Mean| & \rmfamily 0.1778s & \rmfamily  0.1240s & \rmfamily  \textcolor{green}{\textbf{30.25\%}} Faster \\
  \verb|Lower Quartile| & \rmfamily 0.1328s & \rmfamily  0.089s & \rmfamily  \textcolor{green}{\textbf{32.96\%}} Faster \\
  \verb|Median| & \rmfamily 0.163s & \rmfamily  0.1255s & \rmfamily  \textcolor{green}{\textbf{23.01\%}} Faster\\
  \verb|Upper Quartile| & \rmfamily 0.226s & \rmfamily  0.163s & \rmfamily  \textcolor{green}{\textbf{27.87\%}} Faster \\
  \bottomrule[1.5pt]
\end{tabular}
\end{center}

\textbf{Trivariate  polynomials}
On average only 7\% faster; one example 28\% slower.

\begin{center}
\begin{tabular}{@{}*4l@{}}
  \toprule[1.5pt]
  \multicolumn{1}{c}{\head{Lift}} &
    \multicolumn{2}{c}{\head{Results}} \\ 
  \head{} & \head{Classical} & \head{Incremental} & \head{}\\
  \cmidrule(l){1-1}\cmidrule(r){2-4}
  \verb|Variance| & \rmfamily 0.05541s & \rmfamily  0.06838s & \rmfamily  \textcolor{red}{\textbf{18.96\%}} Larger\\
  \verb|Mean| & \rmfamily 0.2880s & \rmfamily  0.2687s & \rmfamily  \textcolor{green}{\textbf{6.707\%}} Faster \\
  \verb|Lower Quartile| & \rmfamily 0.1275s & \rmfamily  0.0995s & \rmfamily  \textcolor{green}{\textbf{21.96\%}} Faster \\
  \verb|Median| & \rmfamily 0.207s & \rmfamily  0.164s & \rmfamily  \textcolor{green}{\textbf{20.77\%}} Faster\\
  \verb|Upper Quartile| & \rmfamily 0.3605s & \rmfamily 0.3523s & \rmfamily  \textcolor{green}{\textbf{2.29\%}} Faster \\
  \bottomrule[1.5pt]
\end{tabular}
\end{center}

So the lifting code shows opposite results to projection with the savings decreasing with input size: indicating that the overheads required for the incremental work grow faster than the savings (at least for the size of examples studied). 

Of course, we can also experiment with combined projection and lifting.  Unsurprisingly the lifting costs dominate.  For the bivariate polynomials incremental code was 37\% faster but for the trivariate only 12\% faster (see \cite{SummerReport} for details).   We think the reason for such drops in performance was due to poor choices of \textsc{Maple}'s data-structure: in particular \textsc{Maple} lists which are implemented as immutable types meaning our edits of them caused separate lists to be created each time. Further, the project may benefit from a fully object-oriented approach.  It is likely further progress could come through code re-factoring.

\section{Summary and Future Work}

The presented work acts as a proof of concept that incremental CAD construction in \textsc{Maple} is possible with savings on offer.  Care needs to be given to the datatypes used.  Beyond that the main areas of further work are moving out of the open case (our implementation restricted lifting to open cells), considering what happens if the new polynomial has a variable not already represented in the system, and considering incremental reduction of constraints.  
To remove a polynomial from the CAD means finding all those projection polynomials created from only that source polynomial (or as a resultant of that with another) and removing them and the cylinder splits their real roots caused.

{\small
\subsubsection*{Acknowledgements}
This work was funded by the EU's H2020 programme under grant No H2020-FETOPEN-2015-CSA 712689 (\scsc).
We thank C.~Brown for a tutorial on the Lazard valuation; 
S.~Timms, J.H.~Davenport and S.~Forrest for useful discussions; 
and the organisers of the \scsc 2017 Summer School where these took place.
}

\bibliographystyle{plain}
\bibliography{references}

\begin{algorithm}[H]
\caption{LiftSetupAdd}
\label{alg:10}
{\fontsize{10}{15}\selectfont
\begin{algorithmic}[1]
\STATE \textbf{Input:} A sets of new projection polynomials $new$, an $oldcad$, a table of sets of all  projection polynomials $psetfull$, and a variable ordering
\STATE \textbf{Output:} $[NewCells, OldCad, OldRoots, UnchangedCells]$ where  NewCells are in the last variable 
$LiftIncf_2$ is information for the next lift, OldCad contains the previous CAD tree, and UnchangedCells is a subset.
\STATE \textbf{Procedure} \texttt{LiftSetup}
\STATE $cad \leftarrow table()$
\STATE $NewRoots \leftarrow [\,]$
\STATE $NewCells \leftarrow table()$
\STATE $UnchangedCells \leftarrow table()$
\STATE $LiftIncf_2 \leftarrow [\,]$ 
\FOR{\texttt{$i$ from $1$ to size$(new[1])$}}
\STATE Append to $NewRoots$ output of $RealRoots(new[1][i])$  
\STATE $NewRoots \leftarrow$ Sort in ascending order and remove duplicates
\ENDFOR 
\STATE $NewCells[1],UnchangedCells[1]=
\texttt{Split}(OldRoots,NewRoots,OldCad)$
\FOR{\texttt{$i$ from $1$ to size$(oldcad[1])$}}
\FOR{\texttt{$j$ from $1$ to size$(new[2])$}}
\STATE Add $oldcad[1][i]$  to $NewCells$
\ENDFOR 
\ENDFOR 
\FOR{\texttt{$i$ from $1$ to size$(NewCells[1])$}}
\FOR{\texttt{$j$ from $1$ to size$(psetfull[2])$}}
\STATE Set $roots$ to real roots of \texttt{LazardValuation}$(oldcad[1][i],new[2][j]$
\STATE Append $[[i],[roots]]$ to $LiftIncf_2$
\ENDFOR 
\ENDFOR 
\FOR{\texttt{$i$ from $1$ to size$(oldcad[1])$}}
\STATE \textbf{If} cell $OldCad[1][i]$'s flag is not equal to \emph{new}, then add cell to $UnchangedCells[1]$ and update index accordingly. 
\ENDFOR 
\STATE \textbf{return} list of variables as in Output
\end{algorithmic}
}
\end{algorithm}

\begin{algorithm}
\caption{LiftAdd}
\label{alg:liftadd}
{\fontsize{10}{15}\selectfont
\begin{algorithmic}[1]
\STATE \textbf{Input:} A sets of new projection polynomials $new$, an $oldcad$, a table of sets of all  projection polynomials $psetfull$, and a variable ordering
\STATE \textbf{Output:} Incremented CAD
\STATE \textbf{Procedure} \texttt{LiftAdd}
\STATE $[NewCells, OldCad, LiftIncf_2, Unchanged] \leftarrow \texttt{LiftSetupAdd}(pset,vars)$
\STATE $dim \leftarrow$ Number of elements in $vars$
\FOR{\texttt{$d$ from $2$ to $dim-1$}} 
\STATE $LiftIncf_{d+1}\leftarrow [\,]$ 
\STATE $NewCells[d] \leftarrow Lift(NewCells_{d-1},LiftIncf_d$ ) 
\FOR{\texttt{$i$ from $1$ to size$(OldCad[d])$}}
\FOR{\texttt{$j$ from $1$ to size$(new[d+1])$}}
\STATE Add $oldcad[d][i]$  to $NewCells$
\ENDFOR 
\ENDFOR 
\FOR{\texttt{$i$ from $1$ to size$(NewCells[d])$}}
\FOR{\texttt{$j$ from $1$ to size$(psetfull[d+1])$}}
\STATE Set $roots$ to real roots of \texttt{LazardValuation}$(OldCad[d][i],new[d+1][j]$
\STATE Append $[[i],[roots]]$ to $LiftIncf_{d+1}$
\ENDFOR 
\ENDFOR 
\FOR{\texttt{$i$ from $1$ to size$(OldCad[d])$}}
\STATE \textbf{If} cell $OldCad[d][i]$'s flag is not equal to \emph{new}, then add cell to $Unchanged[d]$ and update index accordingly. 
\ENDFOR 
\ENDFOR 
\STATE $FinalUnchangedCells \leftarrow []$
\STATE $FinalUnchangedCells$ Union $Unchanged[d][f]$ for all cells $f$ with their corresponding flags not equal to \emph{new}. 
\STATE $IncCAD \leftarrow FinalUnchangedCells$ Union $NewCells[d]$ 
\STATE \textbf{return} IncCAD
\end{algorithmic}
}
\end{algorithm}

\end{document}